\documentclass[twocolumn,prl, reprint, floatfix]{revtex4-1}
\usepackage{hyperref}
\usepackage{units}
\usepackage{amsthm}
\usepackage{amsmath}
\usepackage{setspace}
\usepackage{amssymb}
\usepackage[latin1]{inputenc}
\usepackage[T1]{fontenc}
\usepackage[dvips]{graphicx}
\usepackage{color}
\usepackage[UKenglish]{babel}
\usepackage{subfigure}
\usepackage{ulem}

\newcommand{\RANGLE}%
{\mathchoice{\bigr\rangle}{\bigr\rangle}{\rangle}{\rangle}}
\newcommand{\LANGLE}%
{\mathchoice{\bigl\langle}{\bigl\langle}{\langle}{\langle}}

\renewcommand{\emph}[1]{\textit{#1}}

\begin{document}

\title{Transition between mechanisms of laser-induced field-free molecular orientation}

\author{I. Znakovskaya$^{1}$, M. Spanner$^{2}$, S. De$^{3,4}$, H. Li$^3$, D. Ray$^3$, P. Corkum$^{2,5}$, I.V. Litvinyuk$^{3,6}$, C.L. Cocke$^3$, M.F. Kling$^{1,3}$}

\affiliation{$^1$Max-Planck Institute of Quantum Optics, Hans-Kopfermann-Str. 1, D-85748 Garching, Germany\\$^2$Steacie Institute for Molecular Sciences, National Research Council of Canada, Ottawa, ON, Canada K1A 0R6\\$^3$J.R. Macdonald Laboratory, Physics Department, Kansas State University, 116 Cardwell Hall, Manhattan, KS 66506, USA\\$^4$Saha Institute of Nuclear Physics, 1/AF Bidhannagar, Kolkata 700064, India\\$^5$Joint Attosecond Science Laboratory, University of Ottawa and National Research Council of Canada, 100 Sussex Drive, Ottawa, ON, Canada\\$^6$Centre for Quantum Dynamics and Australian Attosecond Science Facility, Griffith University, Nathan, Queensland, Australia 4111}

\begin{abstract}
The transition between two distinct mechanisms for the laser-induced field-free
orientation of CO molecules is observed via measurements of orientation revival
times and subsequent comparison to theoretical calculations.  In the first mechanism,
which we find responsible for the orientation of CO up to peak intensities of
$8\times10^{13}$ W/cm$^2$, the molecules are impulsively oriented through the
hyperpolarizability interaction. At higher intensities, asymmetric depletion
through orientation-selective ionization is the dominant orienting mechanism.
In addition to the clear identification of the two regimes of orientation, we
propose that careful measurements of the onset of the orientation depletion
mechanism as a function of the laser intensity will provide a relatively simple
route to calibrate absolute rates of non-perturbative strong-field molecular
ionization.
\end{abstract}

\maketitle

%%%%%%%%%%%%%%%%%%%%%%%%%%%%%%%%%%%%%%%%%%%%%%%%%%%%%%%%%%%%%%%%%%%%%%%%%%%%
% Intro
%%%%%%%%%%%%%%%%%%%%%%%%%%%%%%%%%%%%%%%%%%%%%%%%%%%%%%%%%%%%%%%%%%%%%%%%%%%%

Laser-induced field-free molecular alignment has become a routine tool in
studies of ultrafast dynamics of small molecules \cite{TamarReview}, ranging
from experiments on attosecond dynamics \cite{AttoReview} and high-harmonic
generation \cite{Sakai2005} to the investigations of coupled
electronic-vibrational dynamics \cite{AlbertCS2}.  In this technique, a strong
but non-ionizing ultrafast laser field is used to give an impulse torque to the
molecules via the polarizability interaction.  After the pulse is over, the
dominant axis of polarizability of molecules briefly aligns along the
polarization direction of the laser as the molecules undergo quantum rotational
revival dynamics, thereby effectively allowing experiments to be carried out in
the molecular frame.  However, for all its success, laser-induced alignment
does not differentiate between the two different polarities of polar molecules.
Thus, when applied to polar molecules, for example CO, the bond axis can be
aligned in space but the direction in which the C or O ends point is not
controlled.  Achieving control over this latter property, called molecular
orientation, has proved to be a most challenging experimental task.
Field-free orientation was initially generated using a combination of lasers
and static electric fields \cite{Sakai2008,Ghafur2009,Holmegaard2010}.  It was
only recently that an all-optical, and hence simpler, technique for
laser-induced field-free molecular orientation had been demonstrated first in
Refs. \cite{De2009,Sakai2010} followed by Ref.\cite{Frumker2012}.  Although
these experiments all relied on the use of a two-color pump pulse comprised of
the fundamental frequency $\omega$ and its second harmonic $2\omega$ to orient
ensembles of molecules, they offered differing interpretations for the underlying
mechanism leading to orientation.  The first two studies
\cite{De2009,Sakai2010} attributed the orientation to the hyperpolarizability
(HP) interaction \cite{Kanai2001}, while the third \cite{Frumker2012} claimed
that an ionization depletion (ID) mechanism \cite{Spanner2012} was active.

Here we report on the experimental observation of a transition between the two
proposed mechanisms, thereby connecting the two differing interpretations
offered in Refs.\cite{De2009,Sakai2010} and \cite{Frumker2012}.  We present
intensity-dependent measurements of the laser-induced orientation of CO and
compare them to recent theoretical predictions by Spanner \textit{et al},
considering both the HP and the ID mechanisms \cite{Spanner2012}.  By comparing
the measured and calculated temporal structures of the revivals of orientation,
we unambiguously assign the regimes of the HP and ID mechanisms.  This
assignment based on the temporal structure of the revivals is further supported
by the intensity dependence of the maximum orientation, which displays a sharp
change in slope as the ID mechanism becomes active.  For CO molecules, we find
that at intensities below $8\times10^{13}$ W/cm$^2$ the HP mechanism is
responsible for orientation, while the ID mechanism becomes dominant at higher
intensities.  In addition, we find that the ID mechanism is responsible for
generating the highest degrees of orientation we observe.

\begin{figure*}[htp]
\onecolumngrid
\centering
\includegraphics[width=\textwidth,keepaspectratio=true,clip=true,trim=0mm 0mm 0mm 0mm]{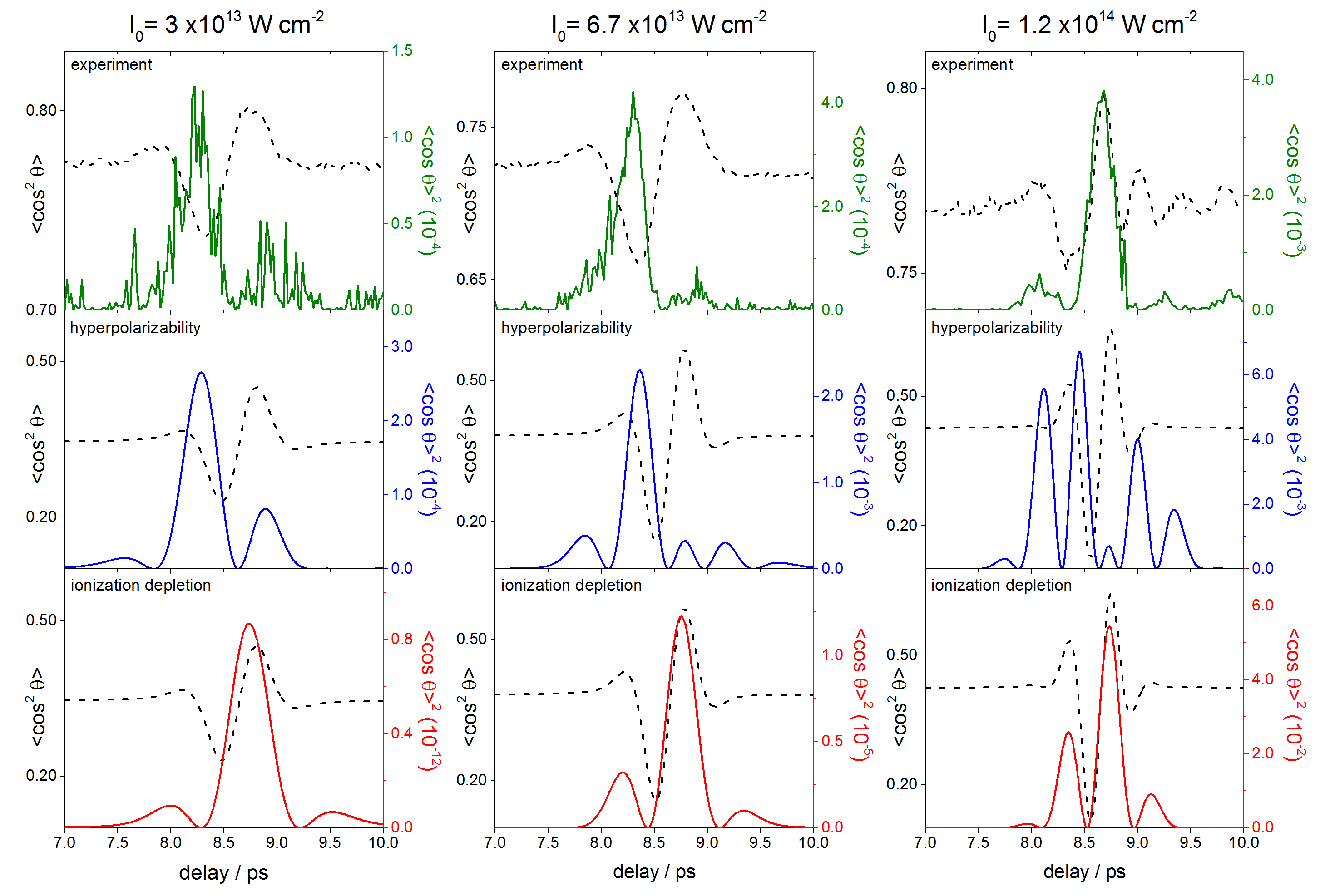}
\caption{(Color online) Comparison of experimental dynamic alignment and
orientation data for CO at three selected intensities of the two-color field
(top row) with theoretical predictions from the hyperpolarizability (middle
row) and ionization depletion mechanisms (bottom row). The alignment (dashed
black lines in all rows) is represented by the parameter $\langle\cos^2\theta\rangle$.
The orientation is given by the parameter $\langle \cos\theta\rangle^2$.}
\label{fig_delay}
\twocolumngrid
\end{figure*}

% Using this method, a maximum orientation of
%$\langle \cos \theta_{\text{exp}}\rangle$ = 0.06 was reported, which was in
%reasonable agreement at the time with theoretical predictions based on the HP
%mechanism.  In Ref.\cite{Frumker2012}, the probe pulse instead was used to
%induce high-harmonic generation (HHG) of the dynamically-evolving ensemble of
%oriented neutral molecules. In this case, the orientation breaks the symmetry
%in HHG and results in even harmonic emission.  The ratio between even and odd
%harmonic intensity was used to estimate the degree of orientation.  A higher
%value of $\langle \cos\theta\rangle$ = 0.2 was reported, and based on the
%temporal structure of the orientation revivals it was argued that the
%orientation was a consequence of the ID mechanism.

%%%%%%%%%%%%%%%%%%%%%%%%%%%%%%%%%%%%%%%%%%%%%%%%%%%%%%%%%%%%%%%%%%%%%%%%%%%%
% Experiment
%%%%%%%%%%%%%%%%%%%%%%%%%%%%%%%%%%%%%%%%%%%%%%%%%%%%%%%%%%%%%%%%%%%%%%%%%%%%

The two-color method of laser-induced orientation uses a pump pulse created by
combining a linearly polarized pulse at its fundamental ($\omega$) and second
harmonic frequency (2$\omega$). The symmetry along the polarization axis of the
resulting field $E(t)=E_{\omega}(t) \cos(\omega t) + E_{2 \omega}(t) \cos(2
\omega t + \varphi)$ depends on the phase delay $\varphi$ between the two
frequency components.  Breaking this symmetry is the key to the orientation of
heteronuclear molecules.  The experimental setup is detailed in \cite{De2009}.
Briefly, linearly polarized pulses with 45~fs duration at 800~nm are produced
using a Ti:sapphire laser and are split into a pump and a probe arm of a
Mach-Zehnder interferometer.  In the pump arm, the second harmonic is created,
temporally synchronized and rotated in its polarization with respect to the
fundamental. A rotatable calcite plate serves to adjust the relative phase
$\varphi$ between the two-colors of the excitation field. The phase is
calibrated by measurement of the phase-dependent ionization of CO
\cite{Li2011}. Note, that this method corrects an error of $\pi$ in the
assignment of $\varphi$ reported in De \textit{et al}., which was based on
above-threshold ionization \cite{De2009}.  The resulting field-asymmetric
two-color pump pulses are focused onto a supersonic jet of CO molecules
(T$_{rot} \approx$ 60 K) inside a velocity-map imaging spectrometer (VMIS)
\cite{Parker1997} by a spherical mirror (f~=~75~mm) placed at the rear side of
the VMIS.  The intensity of the two-color pump pulses was varied between
2.4$\times$10$^{13}$ W/cm$^2$ and 1.2$\times$10$^{14}$ W/cm$^2$.

The degree of orientation and alignment induced by the pump pulse is typically
characterized by the observables $\langle \cos\theta\rangle$ and $\langle
\cos^2\theta\rangle$ respectively, where $\theta$ is the angle between the
molecular axis and the polarization direction of the laser. It is not trivial
to directly measure the angle $\theta$.  As in Ref.\cite{De2009}, we use a
strong single-color (800 nm) probe pulse to multiply ionize and Coulomb explode
the molecule.  The angle $\theta$ was then approximated by the angle $\theta
\approx \theta_{\text{exp}}$ at which fragments arising from the Coulomb
explosion are detected with the VMIS.  The intensity of the probe pulse is
(2.6$\pm$0.6)$\times$10$^{14}$ W/cm$^2$.  Similar to the studies reported in
Ref.\cite{De2009}, we have chosen to analyze the angular emission of C$^{2+}$
fragments at kinetic energies above 2.5~eV. The top row of
Fig.\,\ref{fig_delay} shows the experimental alignment traces (dashed black
lines) around the first full revival of $\langle \cos^2 \theta\rangle$ near the
molecules' full rotation time 1/(2Bc) of 8.64 ps (with B = 1.93 cm$^{-1}$)
\cite{NIST} at three selected two-color pump intensities. Also shown are the
corresponding experimental orientation traces (green lines), which are
reflected here by the parameter $\langle \cos\theta\rangle^2$, following the
presentation in \cite{Spanner2012}.  Plotting the square of the orientation
makes it easier to compare the maximum of the orientation relative to the
alignment revival signal.  We note that the linearly polarized probe pulse
increases the measured degree of alignment.  Furthermore, since the probability
of ionizing to the C$^{2+}$ fragment as a function of the actual $\theta$ is
not rigorously known, both the alignment and orientation values extracted from
the Coulomb explosion will be proportional to, but not exact representations
of, the $\langle \cos\theta\rangle$ and $\langle \cos^2\theta\rangle$
observables.

%%%%%%%%%%%%%%%%%%%%%%%%%%%%%%%%%%%%%%%%%%%%%%%%%%%%%%%%%%%%%%%%%%%%%%%%%%%%
% Theory
%%%%%%%%%%%%%%%%%%%%%%%%%%%%%%%%%%%%%%%%%%%%%%%%%%%%%%%%%%%%%%%%%%%%%%%%%%%%

We now turn to the theoretical description of the orientation mechanisms.
For the computations, the two-color laser pulse is written as
\begin{equation}\label{EqEt}
	E(t) = E_0 f(t) [\cos(\omega t) + \cos(2\omega t) ],
\end{equation}
where $f(t)$ is the pulse envelope, $E_0$ is the peak electric field strength.
In this definition of the $E(t)$, we have set the relative phase between the
colors to zero, $\varphi=0$, which assumes the maximum field asymmetry.
We choose the envelope function $f(t)$ to be
\begin{equation}
	f(t) = \left \{
	\begin{array}{l l}
		0, & t<0 \\
		\sin\left({\pi t}/{2\tau_{on}}\right), & 0<t<2\tau_{on} \\
		0, & \:t>2\tau_{on}
	\end{array} \right.
\end{equation}
corresponding to a $\sin^2$ pulse for the intensity $I=E^2$.  The parameter
$\tau_{on}$  is the full width at half-intensity, which we set to
$\tau_{on}=45$ fs.

The rotational motion of the molecules does not follow the carrier oscillations
of the laser, and hence it is appropriate to use the cycle-averaged Hamiltonian
of the system to compute the rotational dynamics
\begin{equation}\label{EqHamil}
	H(\theta,t) = BJ(J+1) + V_P(\theta,t) + V_H(\theta,t) + V_I(\theta,t),
\end{equation}
where $B$ is the rotational constant, $V_P(\theta,t)$ is the polarizability
term that generates molecular alignment \cite{Herschbach}, $V_H(\theta,t)$ is
the hyperpolarizability term, and $V_I(\theta,t)$ accounts for ionization
\cite{Spanner2012}.  All equations use Hartree atomic units ($m_e=e=\hbar=1$).
For the particular $E(t)$ chosen in Eq.~\ref{EqEt}, the first two potential
terms in Eq.~\ref{EqHamil} are given by \cite{Kanai2001}
\begin{equation}
	V_P(\theta,t) = -\frac{1}{2} \Delta\alpha E_0^2 |f(t)|^2 \cos^2\theta
\end{equation}
\begin{eqnarray}\label{EqVhyper}
	V_H(\theta,t) &=& -\frac{3}{8}\beta_{xxz}E_0^3|f(t)|^3 \cos\theta \\ \nonumber
	              & & -\frac{1}{8}(\beta_{zzz}-3\beta_{xxz})E_0^3|f(t)|^3\cos^3\theta
\end{eqnarray}
where $\Delta\alpha = \alpha_\parallel - \alpha_\perp$ is the polarizability
anisotropy, and the $\beta_{ijk}$ are elements of the hyperpolarizability
tensor.  The ionization depletion term is given by
\begin{equation}\label{EqSeparableIonization}
	V_I(\theta,t) = -(i/2)K(t) \Gamma_{\rm ref}(\theta),
\end{equation}
where
\begin{equation}\label{EqTunExp}
	K(t) =  \exp\left(-\frac{2}{3}(2I_p)^{3/2}\left[|E_0f(t)|^{-1} - |E_{\rm ref}|^{-1}\right]\right)
\end{equation}
accounts for the tunneling exponent \cite{Keldysh} that provides the
dominant scaling of strong-field ionization, and
\begin{equation}\label{EqItheta}
	\Gamma_{\rm ref}(\theta) = c_0 + c_1\cos\theta + c_2 \cos 2\theta
\end{equation}
accounts for the angle-dependence of the ionization rate.  As outlined in
Ref.\cite{Spanner2012}, this analytical form of $V_I(\theta,t)$ is specific to
CO and was constructed as a fit to purely numerical computations that used the
method of Ref.\cite{SpannerPatchkovskii}.  It is a complex potential causing
non-unitary quantum evolution that removes amplitude as a function of angle,
which captures the effects of population loss due to ionization.  All the
molecular constants that appear in the potentials are collected in Table
\ref{TabConstants}.  Although the Hamiltonian (Eq. \ref{EqHamil}) includes both the
hyperpolarizability and ionization terms, we present  results for each
mechanisms separately in order to clearly elucidate the characteristic features
of the two mechanisms.

\begin{table}[tbc]
    \caption{Molecular constants (a.u.) used to model CO.}
    \begin{ruledtabular}
    \begin{tabular}{cc|cc}
    Parameter & Value [Ref.] & Parameter & Value [Ref.] \\
    \hline
    $B$            & 8.7997$\times 10^{-6}$  \cite{NIST}     &
    $\Delta\alpha$ & 3.6                     \cite{PetersonDunning} \\
    $\beta_{zzz}$  & 28.91                   \cite{PetersonDunning} &
    $\beta_{xxz}$  & 7.69                    \cite{PetersonDunning} \\
    $I_p$          & 0.516                   \cite{NIST}     &
    $c_0$          & 0.2214 $ \times 10^{-3}$  \\
    $E_{\rm ref}$  & 0.0535                    &
    $c_1$          &-0.2141 $ \times 10^{-3}$   \\
	               &                           &
    $c_2$          & 0.0822 $ \times 10^{-3}$
    \end{tabular}
    \end{ruledtabular}
    \label{TabConstants}
\end{table}

The time-dependent rotational Schr\"odinger equation for the Hamiltonian in
Eq.~(\ref{EqHamil}) is solved in a spherical harmonics basis using the
Crank-Nicholson method \cite{NumericalRecipes}.  We account for the thermal
distribution by propagating each initial rotational state $|J,M\rangle$
independently and then incoherently averaging the $\langle \cos^2\theta\rangle$
and $\langle \cos\theta\rangle$ observable weighted by the Boltzmann
distribution at temperature $T=60$ K.  For the ionization depletion mechanism,
the observables are further normalized to the remaining neutral population.

The results of our calculations for the HP (ID) mechanism are shown in the
middle (bottom) row of Fig.\,\ref{fig_delay}, where alignment is displayed as
dashed black lines and orientation as blue (red) lines. It can be seen that the
temporal structure of the measured orientation revival is reflected well by the
HP mechanism for the two lower intensities of 3$\times$10$^{13}$ W/cm$^2$ and
6.7$\times$10$^{13}$ W/cm$^2$.
%Note that the displayed theoretical results have
%not been scaled to match the experimental values.
At these two intensities, the ID mechanism does not yet yield a significant
orientation signal and stays below the expected signal from the HP mechanism.
The situation changes at the highest intensity investigated of
1.2$\times$10$^{14}$ W/cm$^2$. Here, the ID mechanism creates an orientation
that is stronger than the one generated from the HP mechanism. In good
agreement with the theoretical prediction, the experimental orientation trace
resembles the orientation revival predicted by the ID mechanism.  Remarkably,
the transition between the mechanisms is well reflected in a temporal shift of
the dominant orientation peak from around 8.25\,ps and 8.3\,ps at
3$\times$10$^{13}$ W/cm$^2$ and 6.7$\times$10$^{13}$ W/cm$^2$, respectively, to
8.7\,ps at 1.2$\times$10$^{14}$ W/cm$^2$. Close inspection of the timing of the
maximum of the orientation revival is thus a good indicator of the active
orientation mechanism.

%%%%%%%%%%%%%%%%%%%%%%%%%%%%%%%%%%%%%%%%%%%%%%%%%%%%%%%%%%%%%%%%%%%%%%%%%%%%
% Intensity dependence
%%%%%%%%%%%%%%%%%%%%%%%%%%%%%%%%%%%%%%%%%%%%%%%%%%%%%%%%%%%%%%%%%%%%%%%%%%%%

\begin{figure}[t]
\centering
\includegraphics[width=0.5\textwidth,keepaspectratio=true,clip=true,trim=0mm
0mm 0mm 0mm]{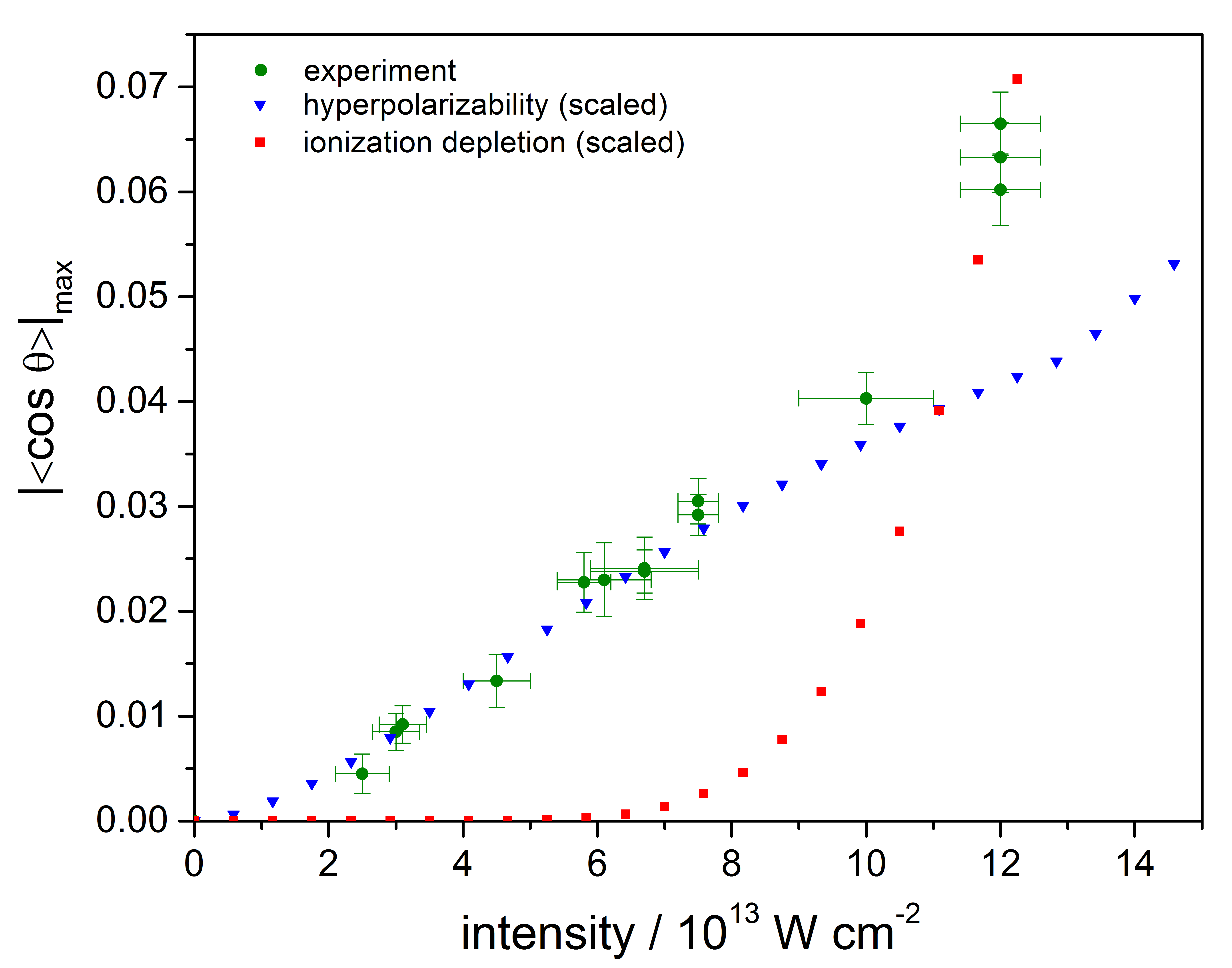} \caption{(Color online) Comparison of the
measured intensity dependence of the maximum field-free orientation
$|\langle \cos\theta\rangle|_{max}$ to predictions for the hyperpolarizability and
ionization depletion mechanisms. The theoretical data for the HP and ID mechanism have been scaled by 0.51 and 0.27, respectively.} \label{fig_intensity}
\end{figure}

As further evidence for the transition between the mechanisms,
Fig.\,\ref{fig_intensity} shows the measured maximum orientation values
$|\langle\cos\theta\rangle|_\text{max}$ (green dots) as a function of the
two-color pump intensity. The horizontal error bars reflect the uncertainty in
the determination of the pump intensity and the vertical error bars the errors in the measured
orientation values. The experimental data is compared to the predictions from the HP (blue triangles)
and ID (red squares) mechanisms. The theoretical
$|\langle\cos\theta\rangle|_\text{max}$ values are scaled by a constant,
intensity independent factor to give best quantitative agreement with the
experimental data. Since the analysis of the data in Fig.\,\ref{fig_delay}
suggests that the HP mechanism is dominant at low intensities, the theoretical
data for the HP mechanism was scaled to fit best to the data below
8$\times$10$^{13}$ W/cm$^2$.  It is remarkable, that the measured gradient of
$|\langle\cos\theta\rangle|_\text{max}$ with intensity is so well predicted by
the HP mechanism in this regime, lending further support for this assignment to
be correct.  At the intensities above 8$\times$10$^{13}$ W/cm$^2$, the
experimental points diverge sharply from the HP curve and display a rapid
increase.  This rapid increase in the experimental maximum orientation matches
qualitatively the exponential increase we would expect from the theoretical
predictions of ID mechanism.

% Based on our assignment, the theoretical data for the ID mechanism
% was scaled to fit the data at the highest intensity of 1.2$\times$10$^{14}$
% W/cm$^2$. Fig.\,\ref{fig_intensity} indicates that ionization depletion can
% indeed lead to a large increase in orientation.

% In this mechanism,  the orientation parameter is expected to increase highly
% non-linearly as a function of increasing pump intensity, and thus appears
% better suited for achieving high degrees of orientation.  Its strength
% depends on an angular asymmetry of the ionization probability along the field
% axis.

The transition between the HP and ID mechanisms offer a unique opportunity for
the calibration of strong-field ionization (SFI) rates.  At present, there is
no reliable methods of computing absolute SFI rates for molecules.  Many
approaches exist to calculate the molecular SFI, from tunneling and
semiclassical models like MO-SFA \cite{Becker} and MO-ADK \cite{Tong2002}, to
more numerically intensive methods like TD-DFT \cite{TDDFT}, TD-CIS
\cite{TDCIS}, MCTDHF \cite{MCTDHF}, and the mixed orbital-grid method of
Ref.\cite{SpannerPatchkovskii}.  These methods have certainly provided much
insight into molecular SFI, and can explain reasonably well the general
intensity scaling and angular dependence of  molecular SFI.  However, the
ultimate reliability of these theories with respect to absolute ionization
rates has not been tested.  Although attempts to measure the absolute SFI rates of
molecules can be found \cite{Alnaser2004,Smeek2011}, they are by no means routine.
Detailed and well calibrated measurements of the transition between the HP and
ID mechanisms provide a means to calibrate the strongly-non-perturbative SFI
response of polar molecules against the perturbative (and hence much better
understood) hyperpolarizability response in one clean experiment.  One can
first ensure the modelling reproduces the orientation in the HP regime to
yield a properly calibrated measurement, than extract the absolute SFI rate
fitting the orientation in the ID regime.  Since the uncharacterized angular
probability of Coulomb exploding to the C$^{2+}$ charge state presently
prevents a quantitative extraction of SFI rate, future work would need to focus
on finding a more reliable experimental measure of
$|\langle\cos\theta\rangle|_\text{max}$.

% Despite considerable progress in the development of theories to treat
% strong-field molecular photoionization, including MO-ADK, SFA or other
% approaches \textbf{[references]}, quantitative prediction of molecular
% photoionization rates remains difficult. While the intensity-dependence of the
% photoionization rate can be reasonably predicted, the absolute rates are often
% off by an order of magnitude or more \textbf{[references]}. The comparison of
% measured degrees of orientation in the ID regime with theoretical calculations
% provides a potential rigid test for theories and may even give access to
% absolute photoionization rates for heteronuclear molecules.

%%%%%%%%%%%%%%%%%%%%%%%%%%%%%%%%%%%%%%%%%%%%%%%%%%%%%%%%%%%%%%%%%%%%%%%%%%%%
% Conclusion
%%%%%%%%%%%%%%%%%%%%%%%%%%%%%%%%%%%%%%%%%%%%%%%%%%%%%%%%%%%%%%%%%%%%%%%%%%%%

In conclusion, we have studied the intensity dependent transition between two
mechanisms for the orientation of CO molecules in two-color laser fields. At
low intensities the HP mechanism is active, while at higher intensities the ID
mechanism dominates.  Apart from reconciling the two differing mechanisms
proposed in the literature, this assignment is important for the future
applicability of the two-color orientation technique.  Since the highest
degrees of orientation are reached via the ID mechanism, achieving large
orientation will always entail the generation of large numbers of cations and
free electrons in the sample.  This necessary generation of cations and
electrons that comes with large orientation could affect experiments using the
two-color orientation technique, and it may be important to take them into
account when using the technique as a tool in subsequent experiments.  In
addition to resolving the active mechanism in laser-induced molecular
orientation, the observation of the transition between the two mechanism offers
a potential experimental observable that can be used to calibrate or extract
the absolute strong-field ionization rates of molecules, which is presently a
remaining computational challenge for all existing theories of strong-field
ionization.

%\begin{acknowledgements}
We are grateful for fruitful discussions with E. Frumker. We acknowledge
support by the U.S. Department of Energy under DE-FG02-86ER13491, the DFG via
Kl-1439/3, and Kl-1439/5, and the Cluster of Excellence: Munich Center for
Advanced Photonics (MAP).
%\end{acknowledgements}

\bibliographystyle{apsrev}
\bibliography{Literature}

\end{document}